\documentstyle[aps,preprint,epsf]{revtex}
\newcommand{\beq}{\begin{equation}}
\newcommand{\eeq}{\end{equation}}
\begin{document}
\draft
\tightenlines

\title{ Appearance of Fermion Condensation Quantum Phase Transition
in Different Fermi Liquids}

\author{ V.R. Shaginyan \footnote{E--mail:
vrshag@thd.pnpi.spb.ru}}
\address{ Petersburg Nuclear Physics
Institute, Russian Academy of Sciences, Gatchina, 188300, Russia}
\maketitle

\begin{abstract}
We show that the quasiparticle effective mass $M^*$
diverges as a function of the system's density $x$,
$M^*\propto 1/(x-x_{FC})$, when a system approaches the
critical point $x_{FC}$ at which the fermion condensation
quantum phase transition (FCQPT) occurs. Such behavior is
of general form and takes place in both three dimensional
systems and two dimensional ones. We demonstrate that a
system which has undergone FCQPT and
lies close to the critical point $x_{FC}$ can be driven
back into the normal Fermi liquid by applying a small magnetic field
$B$. As the field $B$ is reduced, the system is tuned back to the
critical point and the effective mass diverges as  $M^*\propto
1/\sqrt{B-B_{FC}}$ where $B_{FC}$ is the maximum field at which FCQPT
takes place. If the system lies precisely at the critical point then
$B_{FC}=0$.  We demonstrate the constancy of the Kadowaki-Woods ratio
when approaching the critical point at the fields $B\geq B_{FC}$.
Analyzing recent experimental data on the effective mass behavior in
a strongly correlated two dimensional fluid $^3$He, in metallic two
dimensional electron systems and in heavy-fermion systems, we show
that the observed behavior is in agreement with our consideration. As
a result, we may conclude that FCQPT can be conceived of as a
universal cause of the strongly correlated regime in different Fermi
liquids.

\end{abstract}\bigskip

It is widely believed that unusual properties of the high-temperature
superconductors is defined by a quantum phase transition which takes
place at temperature $T=0$ being driven by a control parameter other
then temperature, for instance, pressure or the density of mobile
charge carriers $x$. The quantum phase transition occurs at the
quantum critical point; in a common case, this point is the end of a
line of continuous transitions at $T=0$. As any phase transition, the
quantum phase transition is related to the order parameter which
induces a broken symmetry. By now, experimental data show that there
are no any broken rotational symmetry or broken translational
symmetry, see e.g. \cite{vn}. Therefore, the relevant quantum phase
transition and its quantum critical point have to possess peculiar
features, immediate experimental study of which is of crucial
importance for understanding the physics of the high-temperature
superconductivity and strongly correlated systems.  Unfortunately,
it is difficult to study the properties of the high-temperature
superconductors, which bear a direct relation to the critical point,
because all the corresponding area is occupied by the
superconductivity. On the other hand, recent experimental data on
different Fermi liquids in the highly correlated regime at the
critical point can help to illuminate both the nature of this point
and the control parameter by which this point is driven. Experimental
facts on strongly correlated high-density two dimensional (2D) $^3$He
\cite{mor,cas} show that the effective mass diverges when the density
at which 2D $^3$He liquid begins to solidify is approached
\cite{cas}. Then, sharp increase of the effective mass when the
density tends to the critical density of the metal-insulator
transition point, which occurs at sufficiently low densities, in a
metallic 2D electron system was observed \cite{skdk}. Note, that
there is no ferromagnetic instability in both Fermi systems and the
relevant Landau amplitude $F^a_0>-1$ \cite{cas,skdk}, in accordance
with the almost localized fermion model \cite{pfw}. At $T\to
0$, another critical point in heavy-fermion metal YbRh$_2$Si$_2$ is
observed \cite{gen}. This critical point is driven by magnetic fields
$B$ which suppresses the antiferromagnetic order, when $B$ reaches the
critical value, $B=B_c$, while the effective mass $M^*$ diverges as
$M^*\propto 1/\sqrt{B-B_c}$ \cite{gen}. The study of the magnetic
field dependence of the coefficients $A$, $\gamma_0$, and $\chi_0$ in
the resistivity, $\Delta\rho=A(B)T^2$, specific heat,
$C/T=\gamma_0(B)$, and the magnetic ac susceptibility,
$\chi_{ac}=\chi_0(B)$, has revealed that YbRh$_2$Si$_2$ behaves as a
true Landau Fermi liquid for $B>B_c$ and the well-known
Kadowaki-Woods ratio $A/\gamma_0$ \cite{kadw} is preserved \cite{gen}.
It is pertinent to note that heavy fermion metals are more likely to
be three dimensional (3D) then 2D and nonetheless we observe the same
critical point as in the mentioned above strongly correlated Fermi
liquids.

In this Report, to study the nature of the discussed critical
points, we analyze the appearance of the fermion condensation phase
transition (FCQPT) \cite{ms} in different 2D and 3D Fermi liquids. We
show that at $T\to 0$ FCQPT manifests itself in the divergence of the
quasiparticle effective mass $M^*$ as the density $x$ of a system
approaches the critical point $x_{FC}$ at which FCQPT takes place, so
that $M^*\propto 1/|x-x_{FC}|$. If a Fermi system lies beyond
FCQPT point and the density is close to the critical point, then
FCQPT can be driven by magnetic fields, while the effective mass
diverges as $M^*\propto1/\sqrt{B-B_{FC}}$ where $B_{FC}$ is the
maximum field at which the FCQPT still occurs. As a result, we show
that FCQPT can be conceived of as a universal cause of the
strongly correlated regime in different Fermi liquids.

We start with a brief consideration of general properties of FCQPT
taking, as a case in point, a two-dimensional electron liquid in the
superconducting state, when the system has undergone FCQPT
\cite{ms,ams}.  At $T=0$, the ground state energy $E_{gs}[\kappa({\bf
p}),n({\bf p})]$ is a functional of the order parameter of the
superconducting state $\kappa({\bf p})$ and of the quasiparticle
occupation numbers $n({\bf p})$ and is determined by the known
equation of the weak-coupling theory of superconductivity (see e.g.
\cite{til}) \beq E_{gs}=E[n({\bf p})] +\int \lambda_0V({\bf p}_1,{\bf
p}_2)\kappa({\bf p}_1) \kappa^*({\bf p}_2) \frac{d{\bf p}_1d{\bf
p}_2}{(2\pi)^4}.\eeq Here  $E[n({\bf p})]$ is the ground-state energy
of normal Fermi liquid, $n({\bf p})=v^2({\bf
p})$ and $\kappa({\bf p})=v({\bf p})\sqrt{1-v^2({\bf p})}$.
It is assumed that the pairing interaction
$\lambda_0V({\bf p}_1,{\bf p}_2)$ is weak. Minimizing $E_{gs}$ with
respect to $\kappa({\bf p})$
we obtain the equation connecting the single-particle energy
$\varepsilon({\bf p})$ to the superconducting gap $\Delta({\bf p})$
\beq \varepsilon({\bf p})-\mu=\Delta({\bf p})
\frac{1-2v^2({\bf p})} {2\kappa({\bf p})},\eeq
here $\mu$ is the chemical potential.
The single-particle energy $\varepsilon({\bf p})$ is
determined by the Landau equation \cite{lan}
\beq\varepsilon({\bf p})=\frac{\delta
E[n({\bf p})]}{\delta n({\bf p})}.\eeq
The equation for the
superconducting gap $\Delta({\bf p})$ takes form
\beq \Delta({\bf p})
=-\lambda_0\int V({\bf p},{\bf p}_1)\kappa({\bf p}_1)
\frac{d{\bf p}_1}{4\pi^2}.\eeq
If $\lambda_0\to 0$, then, the maximum value of the superconducting
gap $\Delta_1\to 0$, and Eq. (2) reduces to that proposed
in \cite{ks} \beq \varepsilon({\bf p})-\mu=0,\: {\mathrm {if}}\,\,\,
\kappa({\bf p})\neq 0,\,\, (0<n({\bf p})<1);
\,\,\: p_i\leq p\leq p_f\in L_{FC}.\eeq At $T=0$, Eq. (5)
defines a new state of Fermi liquid with the fermion condensate (FC)
for which the modulus of the order parameter $|\kappa({\bf p})|$ has
finite values in $L_{FC}$ range of momenta $p_i\leq p\leq p_f$
occupied by FC, while the superconducting gap can be
infinitely small, $\Delta_1\to 0$ in $L_{FC}$
\cite{ms,ks,vol}. Such a state can be considered as
superconducting, with infinitely small value of $\Delta_1$ so that
the entropy of this state is equal to zero. This
state, created by the quantum phase transition, disappears at
$T>0$. FCQPT can be considered as a ``pure'' quantum phase transition
because it cannot take place at finite temperatures.
Therefore, its quantum critical point does not represent
the end of a line of continuous transitions at $T=0$.
Equation (5) determines also the critical point of FCQPT, possessing
solutions at some density $x=x_{FC}$.  Nonetheless, FCQPT has a
strong impact on the system's properties up to
temperature $T_f$ above which FC effects become insignificant
\cite{ms,ks}. FCQPT does not violate any rotational symmetry or
translational symmetry, being characterized by the order parameter
$\kappa({\bf p})$. It follows from Eq. (5) that the
quasiparticle system brakes into two quasiparticle subsystems: the
first one in $L_{FC}$ range is occupied by the quasiparticles with
the effective mass $M^*_{FC}\propto 1/\Delta_1$, while the second by
quasiparticles with finite mass $M^*_L$ and momenta $p<p_i$. If
$\lambda_0\neq0$, $\Delta_1$ becomes finite, leading to finite value
of the effective mass $M^*_{FC}$ in $L_{FC}$, which can be obtained
from Eq. (2) \cite{ms,ams} \beq M^*_{FC} \simeq
p_F\frac{p_f-p_i}{2\Delta_1},\eeq while the effective mass $M^*_L$ is
disturbed weakly.  Here $p_F$ is the Fermi momentum.  It follows from
Eq. (6) that the quasiparticle dispersion can be presented by two
straight lines characterized by the effective masses $M^*_{FC}$ and
$M^*_L$ respectively. These lines intersect near the binding energy
$E_0$ of electrons which defines an intrinsic energy scale of the
system: \beq E_0=\varepsilon({\bf p}_f)-\varepsilon({\bf p}_i)
\simeq\frac{(p_f-p_i)p_F}{M^*_{FC}}\simeq 2\Delta_1.\eeq

In fact, as it is seen from Eqs. (5) and (6) even at $T=0$, Fermi
liquid with FC presenting highly generated state is absorbed by the
superconducting phase transition and never exhibits
the dispersionless plateau associated with $M^*_{FC}\to \infty$. As
a result, a Fermi liquid beyond the point of FCQPT can
be described by two types of quasiparticles characterized by the two
finite effective masses $M^*_{FC}$ and $M^*_{L}$ respectively and by
the intrinsic energy scale $E_0$. It is reasonably safe
to suggest that we have come back to the Landau theory by integrating
out high energy degrees of freedom and introducing the
quasiparticles. The sole difference between the Landau Fermi liquid
and Fermi liquid undergone FCQPT is that we have to expand the
number of relevant low energy degrees of freedom by adding both a new
type of quasiparticles with the effective mass $M^*_{FC}$, given by
Eq. (6), and the energy scale $E_0$ given by Eq. (7). We have also to
bear in mind that the properties of these new quasiparticles of a
Fermi liquid with FC cannot be separated from the properties of the
superconducting state, as it follows from Eqs. (6) and (7).
We may say that the quasiparticle system in the range $L_{FC}$
becomes very ``soft'' and is to be considered as a strongly correlated
liquid.  On the other hand, the system's properties and dynamics are
dominated by a strong collective effect having its origin in FCQPT
and determined by the macroscopic number of quasiparticles in the
range $L_{FC}$.  Such a system cannot be disturbed by the scattering
of individual quasiparticles and has features of a quantum
protectorate \cite{ms,lpa}.

Let us assume that FC has just taken place, that is $p_i\to p_f\to
p_F$, the deviation $\delta n(p)$ is small, and $\lambda_0\to 0$.
Expanding functional $E[n(p)]$ in Eq. (3) in Taylor's series with
respect to $\delta n(p)$ and retaining the leading terms, one obtains
taking into account Eq. (5),
\beq \mu=\varepsilon({\bf p},\sigma) =
\varepsilon_0({\bf p},\sigma)+\sum_{\sigma_1}\int F_L({\bf p},{\bf
p}_1,\sigma,\sigma_1)\delta n({\bf p_1},\sigma_1) \frac{d{\bf
p}_1}{(2\pi)^2};\,\,\: p_i\leq p\leq p_f\in L_{FC},\eeq where
$F_L({\bf p},{\bf p}_1,\sigma,\sigma_1)= \delta^2 E/\delta n({\bf
p},\sigma)\delta n({\bf p}_1,\sigma_1)$ is the Landau interaction,
and $\sigma$ denotes the spin states.  Both the Landau interaction
and the single-particle energy $\varepsilon_0(p)$ are calculated at
$n(p)=n_F(p)$. Equation (8) possesses solutions if the
Landau amplitude $F_L$ is positive and sufficiently large, so that
the integral on the right hand side of Eq. (8) defining the potential
energy is large and makes the potential energy prevail over the
kinetic energy $\varepsilon_0({\bf p})$ \cite{ks}. It is also seen
from Eq.  (8) that the FC quasiparticles forms a collective state,
since their energies are defined by the macroscopical number of
quasiparticles within the region $p_i-p_f$, and vice versa. The shape
of the spectra is not effected by the Landau interaction, which,
generally speaking, depends on the system's properties, including the
collective states, impurities, etc. The only thing defined by the
interaction is the width of the region $p_i-p_f$, provided the
interaction is sufficiently strong to produce the FC phase transition
at all.  Thus, we can conclude that the spectra related to FC are of
universal form.  At temperatures $T\geq T_c$, the effective mass
$M^*_{FC}$ related to FC is given by \cite{ms,ams}, \beq
M^*_{FC}\simeq p_F\frac{p_f-p_i}{4T}.\eeq Multiplying both sides of
Eq. (9) by $p_f-p_i$ we obtain the energy scale $E_0$ separating the
slow dispersing low energy part, related to the effective mass
$M^*_{FC}$, from the faster dispersing relatively high energy part,
defined by the effective mass $M^*_{L}$ \cite{ms,ams}, \beq E_0\simeq
4T.\eeq It is seen from Eq. (10) that the scale $E_0$ does not depend
on the condensate volume. The single particle excitations are defined
according to Eq. (9) by the temperature and by $n(p)$, given by Eq.
(5). Thus, again at $T\geq T_c$, the one-electron spectrum is
negligible disturbed by thermal excitations, impurities, etc, and one
observes the features of the quantum protectorate  \cite{ms,lpa}.

FCQPT appears in Fermi liquids, when the effective interaction
becomes sufficiently large, as it occurs in $^3$He at sufficiently
high densities and in many-electron systems at relatively low
density \cite{ksz}. Calculations based on simple models show that FC
can exist as a stable state separated from the normal Fermi liquid
by the phase transition \cite{ks}. In ordinary electron liquid this
interaction is directly proportional to the dimensionless parameter
$r_s\sim 1/p_Fa_B$, where $a_B$ is the Bohr radius. If $p_i\to
p_F\to p_f$, Eq. (5) determines the point $r=r_{FC}$ at which FCQPT
takes place, and $(p_f-p_i)/p_F\sim (r_s-r_{FC})/r_{FC}\sim g_{eff}$,
with $g_{eff}$ being the effective coupling constant,
which characterizes the interaction in proximity to the point
$r_{FC}$ of FCQPT \cite{ms,ks}. FCQPT precedes the formation of
charge-density waves or stripes, which take place at some value
$r_s=r_{cdw}$ with $r_{FC}<r_{cdw}$, while the Wigner solidification
takes place even at larger values of $r_s$ and leads to insulator
\cite{ksz}. On the other hand, there are charge-density waves, or
stripes, in underdoped copper oxides and finally at small doping
levels one has insulators \cite{grun}. In the same way, the effective
mass inevitably diverges as soon as the density $x$ becomes
sufficiently large approaching the critical density at which 2D
$^3$He begins to solidify \cite{ksz}, as it was observed in
\cite{cas}.

Recent studies of photoemission spectra discovered an energy scale
in the spectrum of low-energy electrons in cuprates, which manifests
itself as a kink in the single-particle spectra \cite{vall,blk}. The
spectra in the energy range (-200---0) meV can be described by two
straight lines intersecting at the binding energy $E_0\sim(50-70)$
meV \cite{blk}. Then, in underdoped copper oxides, there are the
pseudogap phenomenon and the line-shape of single-particle
excitations strongly deviates from that of normal Fermi liquid, see,
e.g. \cite{tim}. While, in the highly overdoped regime slight
deviations from the normal Fermi liquid are observed \cite{val1}. All
these peculiar properties are naturally explained within a model
proposed in \cite{ms,ams,ms1} and allow to relate the doping level,
or the charge carriers density $x$, regarded as the density of holes
(or electrons) per unit area to the density of Fermi liquid with FC.
We assume that $x_{FC}$ corresponds to the highly overdoped regime at
which FCQPT takes place. In that case, the effective coupling
constant $g_{eff}\sim (x-x_{FC})/x_{FC}$.  According to the model,
the doping level $x$ at $x\leq x_{FC}$ in metals is related to
$(p_f-p_i)$ in the following way:  \beq g_{eff}\sim
\frac{(x_{FC}-x)}{x_{FC}}\sim \frac{p_f-p_i}{p_F}
\sim \frac{p^2_f-p^2_i}{p^2_F}.\eeq

Now we consider the divergence of the effective mass in 2D and 3D
Fermi liquids at $T=0$, when the density $x$ approaches FCQPT from
the side of normal Landau Fermi liquid (LFL).
First, we calculate the divergence of $M^*$ as a function of the
difference $(x_{FC}-x)$ in case of 2D $^3$He. For this purpose we use
the equation for $M^*$ obtained in \cite{ksz}, where the
divergence of the effective mass $M^*$ due to the onset of FC in
different Fermi liquids including $^3$He was predicted
\beq\frac{1}{M^{*}}=\frac{1}{M}+\frac{1}{4\pi^{2}}
\int\limits_{-1}^{1}\int\limits_0^{g_0}
\frac{v(q(x))}{\left[1-R(q(x),\omega=0,g)\chi_0(q(x)
,\omega=0)\right]^{2}}\frac{xdxdg}{\sqrt{1-x^{2}}}.
\eeq
Here we adopt the shorthand, $p_F\sqrt{2(1-x)}=q(x)$, with $q(x)$
being the transferred momentum, $M$ is the bare mass, $\omega$ is
the frequency, $v(q)$ is the bare interaction, and the integral is
taken over the coupling constant $g$ from zero to its real value
$g_0$. In Eq.  (12), both $\chi_0(q,\omega)$ and $R(q,\omega)$,
being the linear response function of noninteracting Fermi liquid
and the effective interaction respectively, define the linear
response function of the system in question \beq
\chi(q,\omega,g)=\frac{\chi_0(q,\omega)}
{1-R(q,\omega,g)\chi_0(q,\omega)}.
\eeq
In the vicinity of charge density wave instability, occurring at the
density $x_{cdw}$, the singular part of the function $\chi^{-1}$ on
the disordered side is of the well-known form, see.  e.g.  \cite{vn}
\beq\chi^{-1}(q,\omega,g)\propto (x_{cdw}-x)+(q-q_c)^2+(g_0-g),\eeq
where $q_c\sim 2p_F$ is the wavenumber of the charge density wave
order. Upon substituting Eq. (14) into Eq. (12) and taking the
integrals, the equation for the effective mass $M^*$ can be cast
into the following form
\beq\frac{1}{M^*}=
\frac{1}{M}-\frac{C}{\sqrt{x_{cdw}-x}},\eeq with $C$ being some
positive constant. It is seen form Eq. (15) that $M^*$ diverges at
some point $x_{FC}$, which is referred to as the critical point at
which FCQPT occurs, as a function of the difference $(x_{FC}-x)$ \beq
M^*\propto \frac{1}{x_{FC}-x}.\eeq
It follows from the derivation of Eqs. (15) and (16) that the form of
these equations is independent of the bare interaction $v(q)$, therefore
both of these equations are also applicable to 2D electron liquid or
to another Fermi liquid. It is also seen from Eqs. (15) and (16)
that FCQPT precedes the formation of charge-density waves. In
consequence of this, the effective mass diverges at high densities in
case of 2D $^3$He, and it diverges at low density in case of 2D
electron systems, in accordance with experimental facts
\cite{cas,skdk}. Note, that in the both cases the difference
$(x_{FC}-x)$ has to be positive because $x_{FC}$ represents the
solution  of Eq. (15). Thus, considering electron systems we have to
replace $(x_{FC}-x)$ by $(x-x_{FC})$.  In case of 3D system, the
effective mass is given by \cite{ksz}
\beq\frac{1}{M^{*}}=\frac{1}{M}+\frac{p_F}{4\pi^{2}}
\int\limits_{- 1}^{1}\int\limits_0^{g_0}
\frac{v(q(x))xdxdg}{\left[1-R(q(x),\omega=0,g)
\chi_0(q(x),\omega=0)\right]^{2}}.
\eeq
A comparison of Eq. (12) and Eq. (17) shows that there is no
fundamental difference between these equations, and along
the same way we again arrive at Eqs. (15) and (16). The only
difference between 2D electron systems and 3D ones is that FCQPT
occurs at densities which are well below those corresponding to 2D
systems. While in the bulk $^3$He, FCQPT cannot probably take place
being absorbed by the first order solidification.

Consider the divergence of $M^*$ in external magnetic fields $B$,
when an electron system is located very close to the critical point
$x_{FC}$ so that $(x_{FC}-x)/x_{FC}\ll1$. As it
follows from the above, this consideration will be applicable to any
2D or 3D Fermi systems. The application of magnetic field $B$ leads
to a weakly polarized state when some levels at the Fermi level are
occupied by quasiparticles with the ordered spins. These levels are
located in the momentum range $\delta p$ given by \beq \frac{p_F\delta
p}{M^*}\sim B\mu_{eff},\eeq where $\mu_{eff}\sim \mu_B$ is the
effective moment. It is seen from Eq. (8) that this polarized state
lowers the integral on the right hand side of Eq. (8)
by the value $\delta I$,
\beq \delta I\propto \delta p\propto M^*B.\eeq
If $\delta I$ is sufficiently large, Eq. (8) has no solutions
and FC vanishes. As a result, we can conclude that $\delta
I\propto g_{eff}$. Thus, it is seen form Eq. (19) that the strength
of magnetic fields can be used as a control parameter. Let $B_{FC}$
be the magnetic field that suppresses FC so that $p_f\to p_F\to p_I$,
then it follows from Eq.  (11)
\beq (B-B_{FC})M^*\propto g_{eff}\propto
\frac{x-x_{FC}}{x_{FC}}.\eeq
Upon substituting Eq. (20) into Eq. (16) we obtain
\beq M^*(B)\propto \frac{1}{\sqrt{B-B_{FC}}}.\eeq
Equation (21) shows that by applying a magnetic field $B>B_{FC}$,
the system can be driven back into LFL
with the effective mass $M^*(B)$ dependent on the magnetic field.
It was demonstrated that the constancy of the
Kadowaki-Woods ratio is obeyed by systems in the strongly correlated
regime when the effective mass is sufficiently large \cite{ksch}.
Therefore, we are led to the conclusion that by applying magnetic
fields, the system is driven back into LFL, and the constancy of the
Kadowaki-Woods ratio is obeyed. Since the resistivity
$\Delta\rho\propto (M^*)^2$ \cite{ksch}, we obtain from Eq. (21)
\beq \Delta\rho\propto \frac{1}{B-B_{FC}}.\eeq
Obviously, if the system lies precisely at the
critical point, $x=x_{FC}$, then $B_{FC}=0$.
At finite temperatures, there is a temperature
$T^*(B)$ at which the polarized state is destroyed. Then, the value
of the integral in Eq.  (8) is restored, and the system comes back
into the state with $M^*$ defined by Eq. (9) and giving rise to the
non-Fermi liquid behavior (NFL). For instance, in that case
$\Delta\rho\sim T$ \cite{dkss}. At small fields $B$, $T^*(B)$ is an
increasing function of $B$ because $T^*\sim B\mu_{eff}$. Thus, the
higher is the magnetic field $B$, the higher is the temperature
$T^*(B)$, at which the crossover from LFL to NFL occurs.

To explain the nature of the field-induced quantum critical point
in YbRh$_2$Si$_2$, we assume that
the electron system of this heavy fermion metal is located
very close to the critical point $x_{FC}$, so that
$(x_{FC}-x)/x_{FC}\ll1$, and see that the obtained above results are
in good agreement with experimental facts \cite{gen}. We can also
safely assume that the electron system of heavy-fermion metals is 3D
system. Comparatively small values of $T_c$ observed in these metals
rather favor the 3D scenario \cite{mv}. A detailed
analysis of the discussed experimental facts will be published
elsewhere.

To conclude, FCQPT
can be conceived of as a universal cause of the
strongly correlated regime and of non-Fermi liquid behavior in
different Fermi liquids such as 2D $^3$He liquid of high densities,
2D electron systems of low densities, heavy-fermion metals and the
high-temperature superconductors.

This work was supported in
part by the Russian Foundation for Basic Research, No 01-02-17189.


\begin{thebibliography}{99}

\bibitem{vn} C.M. Varma, Z. Nussinov, and Wim van Saarloos, Phys.
Rep. {\bf 361}, 267 (2002).

\bibitem{mor} K.-D. Morhard {\it et al.,} Phys. Rev. B {\bf 53}, 2658
(1996).

\bibitem{cas} A. Casey {\it et al.,} J. Low Temp. Phys. {\bf 113},
293 (1998).

\bibitem{skdk} A.A. Shashkin {\it et al.,} Phys. Rev. B, in press;
cond-mat/0111478.

\bibitem{pfw} M. Pfitzner and P. W\"olfe, Phys. Rev. B {\bf 33}, 2003
(1986).

\bibitem{gen} P. Gegenwart {\it et al.,} Phys. Rev. Lett. {\bf 89},
056402 (2002); P. Gegenwart {\it et al.,} cond-mat/0207570.

\bibitem{kadw} K. Kadowaki and S.B. Woods, Solid State Commun.
{\bf 58}, 507 (1986).

\bibitem{ms} M.Ya. Amusia and V.R. Shaginyan,
JETP Lett. {\bf 73}, 232 (2001);
S.A. Artamonov and V.R. Shaginyan,
JETP {\bf 92}, 287 (2001);
M.Ya. Amusia and V.R. Shaginyan, Phys. Rev. B {\bf 63}, 224507 (2001).

\bibitem{ams} M.Ya. Amusia, S.A. Artamonov, and V.R. Shaginyan,
JETP Lett. {\bf 74}, 435 (2001).

\bibitem{til} D.R. Tilley and J. Tilley, {\it Superfluidity and
Superconductivity}, Bristol, Hilger (1975).

\bibitem{lan} L. D. Landau, Sov. Phys. JETP  {\bf 3}, 920 (1956).

\bibitem{ks} V.A. Khodel and V.R. Shaginyan,
JETP Lett. {\bf 51}, 553 (1990);
V.A. Khodel, V.R. Shaginyan, and V.V. Khodel, Phys. Rep.
{\bf 249}, 1 (1994).

\bibitem{vol} G. E. Volovik, JETP Lett. {\bf 53}, 222 (1991).

\bibitem{lpa} R.B. Laughlin and D. Pines, Proc. Natl. Acad. Sci.
U.S.A. {\bf 97}, 28 (2000); P.W. Anderson, cond-mat/007185;
cond-mat/0007287.

\bibitem{ksz} V.A. Khodel, V.R. Shaginyan, and M.V. Zverev,
JETP Lett. {\bf 65}, 253 (1997).

\bibitem{grun} G. Gr\"{u}ner, {\it Density Waves in Solids,
Addison-Wesley}, Reading, MA (1994).

\bibitem{vall} T. Valla et al., Science {\bf 285}, 2110 (1999).

\bibitem{blk} P.V. Bogdanov {\it et al.,} Phys. Rev. Lett. {\bf 85},
2581 (2000); A. Kaminski {\it et al.,} Phys. Rev. Lett.
{\bf 86}, 1070 (2001).

\bibitem{tim} T.Timusk and B. Statt, Rep. Prog. Phys. {\bf 62}, 61
(1999).

\bibitem{val1} Z. Yusof {\it et al.,} Phys. Rev. Lett.
{\bf 88}, 167006 (2002).

\bibitem{ms1} M.Ya. Amusia and V.R. Shaginyan,
Phys. Lett. A {\bf 298}, 193 (2002).

\bibitem{ksch} V.A. Khodel and P. Schuck, Z. Phys. B {\bf 104}, 505
(1997).

\bibitem{dkss} J.A. Dukelsky {\it et al.,} Z. Phys. B
{\bf 102}, 245 (1997).

\bibitem{mv} M.Ya. Amusia and V.R. Shaginyan, cond-mat/0207306.

\end{thebibliography}
\end{document}